\documentclass[11pt,a4paper]{article}
\pdfoutput=1

\usepackage{jheppub}
\usepackage{cleveref}
\usepackage{subfigure}

\ifx\pdfoutput\undefined
\usepackage[dvips,bookmarks=false]{hyperref}	
\else
\usepackage{hyperref}	
\fi
\hypersetup{colorlinks,bookmarksopen,bookmarksnumbered,citecolor=blue,
linkcolor=black,pdfstartview=FitH,urlcolor=blue}

\makeatletter
\def\@fpheader{Prepared for submission to JCAP}
\makeatother

\newcommand{\be}{\begin{equation}}
\newcommand{\ee}{\end{equation}}

\newcommand{\fPQ}{f_{\rm PQ}}
\newcommand{\PQ}{{\rm PQ}}

\makeatletter

\DeclareRobustCommand{\rcite}[1]{%
  \rcite@aux#1,\@nil{#1}%
}
\def\rcite@aux#1,#2\@nil#3{%
  \if\relax#2\relax
    Ref.~\cite{#3}%
  \else
    Refs.~\cite{#3}%
  \fi
}

\makeatother

\title{Axion minicluster power spectrum and mass function}

\author{Jonas Enander,}
\author{Andreas Pargner,}
\author{Thomas Schwetz}
\affiliation{Institut f\"ur Kernphysik, Karlsruher Institut f\"ur Technologie (KIT),
  76021 Karlsruhe, Germany}
\emailAdd{andreas.pargner@kit.edu}

\abstract{When Peccei-Quinn (PQ) symmetry breaking happens after
  inflation, the axion field takes random values in causally
  disconnected regions. This leads to fluctuations of order one in the
  axion energy density around the QCD epoch. These over-densities
  eventually decouple from the Hubble expansion and form so-called
  miniclusters. We present a semi-analytical method to calculate the
  average axion energy density, as well as the power spectrum, from the
  re-alignment mechanism in this scenario. Furthermore, we develop a
  modified Press \& Schechter approach, suitable to describe the
  collapse of non-linear density fluctuations during radiation
  domination, which is relevant for the formation of axion
  miniclusters. It allows us to calculate the double differential
  distribution of gravitationally collapsed miniclusters as a function
  of their mass and size. For instance, assuming a PQ scale of
  $10^{11}$~GeV, minicluster masses range from about $5 \times
  10^{-16}$ to $3 \times 10^{-13}$ solar masses and have sizes from
  about $4\times 10^4$ to $7\times 10^5$~km at the time they start to
  collapse.}

\begin{document}

\maketitle

\section{Introduction}

The QCD axion \cite{Weinberg:1977ma, Wilczek:1977pj} is one of the
most attractive candidates for the dark matter in the Universe. It is
the Goldstone boson related to a global $U(1)$ symmetry, which is
spontaneously broken at the Peccei-Quinn (PQ) scale
$f_\PQ$~\cite{Peccei:1977ur, Peccei:1977hh}, much larger than the
electro-weak scale. Around the QCD scale the symmetry is explicitly
broken by the potential created by QCD instanton effects.  Axions are
produced by various mechanisms in the early Universe and hence they
can potentially account for the dark matter, see
\rcite{Sikivie:2006ni, Marsh:2015xka} for reviews of axion cosmology.

In this work we are interested in the situation where the PQ phase
transition happens after the end of inflation. In this case the axion
field takes on random values in causally disconnected regions. It was
noted by Hogan and Rees in \rcite{Hogan:1988mp} that this implies
isocurvature fluctuations of order one in the axion energy density at
the QCD phase transition, leading to so-called {\it axion
  miniclusters}. These are gravitationally bound systems of axions,
whose mass is determined roughly by the size of the horizon at the QCD
phase transition. The formation of axion miniclusters has been studied
numerically by Kolb and Tkachev~\cite{Kolb:1993zz,Kolb:1993hw}, and
later by Zurek, Hogan, Quinn~\cite{Zurek:2006sy}. For more recent work
on axion miniclusters see \rcite{Tkachev:2014dpa, Tinyakov:2015cgg,
  Hardy:2016mns, Stadler, Fairbairn:2017dmf,Fairbairn:2017sil}.  Naive
estimates indicate a typical minicluster mass of $10^{-13} M_\odot$,
see e.g.~\rcite{Tinyakov:2015cgg, Davidson:2016uok, Bai:2016wpg} (we are going to
address this number in detail below). If a significant fraction of
axion dark matter is bound in those objects, severe implications for
dark matter axion searches are expected. If dark matter resides in
compact objects of order $10^{-13} M_\odot$, the probability that such
an object passes through a detector at earth is very low, reducing
dramatically the discovery potential of axion haloscopes, see for
instance \rcite{Stern:2016bbw, TheMADMAXWorkingGroup:2016hpc}.
Lensing signals have been discussed in \rcite{Kolb:1995bu}, and
recently in \rcite{Fairbairn:2017dmf,Fairbairn:2017sil}.

In this work we develop semi-analytic methods to understand the
formation of axion miniclusters and their distribution in mass and
size. We restrict our analysis to the axion energy density produced by
the so-called re-alignment mechanism, i.e., a classical, coherently
oscillating axion field. Furthermore, we limit our analysis to the
harmonic approximation of the potential. While those assumptions
clearly capture only part of the full picture, our results do provide
a useful estimate of the properties of the
minicluster distribution generated by the re-alignment mechanism. It
allows to identify important parameters and study in a transparent way
the underlying physics. It will be a useful starting point for
quantitative numerical studies of axion minicluster formation, as well
as the subsequent evolution after decoupling from the Hubble flow. 

The outline of this work is as follows. In \cref{sec:prelim} we set
the stage by reviewing the post-inflationary axion scenario and give a
more concise description of the scope of this work. In
\cref{sec:power} we present a calculation of the average axion energy
density including a consistent treatment of gradient terms, and we
derive the power spectrum of the axion energy density fluctuations.
In \cref{sec:distr} we develop a formalism to describe the
gravitational collapse of the axion over-densities, which are
non-linear from the very beginning. We present the double differential
number density in mass and size of gravitationally bound clumps of
axion energy density around the time of matter-radiation equality.
\Cref{sec:summary} contains the summary and discussion of our results.
In \cref{app:EOM} we describe our method to solve the equation of
motion of the axion field.

\section{Axion preliminaries and goals of this work}
\label{sec:prelim}

The random values of the axion field in causally disconnected regions
in the post-inflation scenario leads to a network of cosmic strings,
with on average one string per Hubble volume. As the Universe expands,
gradient terms will smooth the axion field on scales of the horizon.
The presence of the topological strings plays a crucial role in the
evolution of the massless field \cite{Kibble:1976sj}.  Once the axion
mass turns on due to QCD effects at temperatures around 1~GeV the
string and domain wall network quickly decays \cite{Sikivie:1982qv,
  Davis:1986xc} and will provide a substantial fraction of the energy
density in axions, subject to large uncertainties, see
\rcite{Hiramatsu:2012gg, Kawasaki:2014sqa} for numerical
simulations. In the following we focus on the axions created due to
the re-alignment mechanism (i.e., coherent field oscillations),
neglecting the contribution of strings and domain walls to the average
energy density as well as to inhomogeneities. While keeping in mind
that this can only be part of the real picture, it is still useful to
isolate the contribution of the re-alignment mechanism and study its
properties.

Let us define the dimensionless field describing the axion as
$\theta(\vec{x},t) = A(\vec{x},t) / \fPQ$, where $A(\vec{x},t)$ is the
real scalar field. The equation of motion for $\theta(\vec{x},t)$ in
the expanding Universe is given by
\begin{equation}\label{eq:eom-gen}
  \ddot\theta + 3H(T)\dot\theta - \frac{\nabla^2}{a^2}\theta + V'(\theta,T) = 0 \,.
\end{equation}
Here the dot denotes derivative with respect to time, $\nabla$ is the
derivative with respect to co-moving coordinates, $H(T) = \dot a/a$ is
the expansion rate with the cosmic scale factor $a$, and
$V(\theta,T)$ is the temperature dependent axion potential, and the
prime denotes derivative with respect to $\theta$. The potential is
related to the topological susceptibility of QCD, $\chi(T)$, by
\begin{equation}\label{eq:VT}
  V(\theta,T) = \frac{\chi(T)}{\fPQ^2}(1-\cos\theta) \,.
\end{equation}
For small $\theta$ the cosine can be expanded and we obtain the
temperature dependent axion mass in terms of the susceptibility:
\begin{equation}\label{eq:Vm}
  V(\theta,T) \approx \frac{1}{2}m^2(T)\theta^2 \,,\qquad m^2(T) = \frac{\chi(T)}{\fPQ^2} \,.
\end{equation}
For $T \lesssim 100$~MeV, $\chi(T)$ becomes constant and the axion reaches its zero-temparature mass $m_0$. Approximately we have \cite{Weinberg:1977ma}
\begin{align}
m_0 \simeq \frac{m_\pi f_\pi}{\fPQ}  \frac{\sqrt{m_um_d}}{m_u+m_d}
 \simeq 5.7\times 10^{-6} ~{\rm eV}~ 
\frac{10^{12} \, \rm GeV}{\fPQ} \, ,
\label{eq:mf}  
\end{align}
with $m_\pi$ and $f_\pi$ being the pion mass and decay constant,
respectively, and $m_{u,d}$ are the up, down quark masses.

Below we will allways assume the small $\theta$ expansion. This is a
crucial ingredient of our calculations, since it leads to a linear
equation of motion. It is clear that our results will not include
anharmonic effects when the field takes on values
close to $\theta \simeq \pm\pi$. In the context of miniclusters those
field values may lead to very dense
objects~\cite{Kolb:1993zz, Kolb:1993hw}, which will not be contained in
the mass function derived below and need to be considered as a
correction to our results.

In the harmonic limit the equation of motion for the Fourier modes of
the field decouple:
\begin{equation}\label{eq:eom-k}
  \ddot\theta_k + 3H(T)\dot\theta_k + \omega_k^2 \theta_k = 0 \,,\qquad
\omega_k^2 \equiv \frac{k^2}{a^2} + m(T)^2 \,.
\end{equation}
Qualitatively, we see that super-horizon modes with $\omega_k \ll
3H$ are frozen, $\theta_k = const$, whereas they start to oscillate
once they enter the horizon. We define $T_\text{osc}$ as the
temperature where the zero-mode (i.e., the homogeneous field) starts
to oscillate by the equation
\begin{equation}\label{eq:Tosc}
3H(T_\text{osc}) = m(T_\text{osc}) \,.  
\end{equation}
The corresponding time and scale factors are denoted by $t_\text{osc}$
and $a_\text{osc}$, respectively. Non-zero $k$ modes will start to
oscillate somewhat earlier. The redshifting of non-zero $k$ modes is
encoded by the $1/a^2$ factor in the expression for $\omega_k$ in
\cref{eq:eom-k}. For sufficiently late times the mass term will
dominate for all modes and the energy density will behave like cold
dark matter.

There are two main goals of this work:
\begin{itemize}
\item 
Under the stated assumptions we calculate the energy density in the
axion field based on the solutions of \cref{eq:eom-k}, with initial
conditions motivated by the post-inflation PQ breaking scenario. We
derive an expression for the average axion energy density taking into
account non-zero momentum modes. Furthermore we calculate the power
spectrum of the density fluctuations which eventually will evolve
into the miniclusters.

\item
In order to describe the subsequent evolution of the over-densities
we consider a model of spherical collapse valid during both 
radiation and matter domination. Departing from the energy
density power spectrum, we will apply a modified Press \& Schechter
formalism to estimate the mass and size of the gravitationally bound
clumps of axion dark matter around the time of matter-radiation
equality.
\end{itemize}
Our results will serve as input for the further evolution of the
miniclusters. We do not address the question of how the minicluster
evolves after it has decoupled from the Hubble flow. This is an
important question which, however, is beyond the scope of this work.

For later reference we provide in \cref{tab:indices} a summary of
indicies that we use to denote certain points in the axion field and
minicluster evolution.

\begin{table}
  \centering
  \begin{tabular}{cp{0.8\textwidth}}
    \hline\hline
    Index & Description \\
    \hline
    $i$ & time when we start the field evolution and set the initial conditions for the axion field correlator  with the wave number scale $K \equiv a_i H_i$, the default value is $T_i = 3T_{\rm osc}$ \\
    osc & zero-mode starts to oscillate; defined by $3H(T_\text{osc}) = m(T_\text{osc})$ \\
    1   & quantities at $T=1$~GeV, with $K_1 \equiv a_1H_1$ and $R_1 = 1/K_1$ used as reference scale for wave number or length-scale plots\\
    $\star$ & $T_\star \equiv 100$~MeV, axion mass reaches its zero-temperature value and all $k$ modes are non-relativistic; time when we start spherical collapse\\
    eq  & matter-radiation equality\\
    \hline\hline
  \end{tabular}
  \caption{Summary of indicies to denote quantities like time $t$,
    temperature $T$, Hubble rate $H$, and cosmic scale factor $a$ at
    certain moments of evolution. \label{tab:indices}}
\end{table}

\section{Axion energy density and power spectrum}
\label{sec:power}

\subsection{Initial conditions}
\label{sec:initial}

Let us specify our choice of initial conditions, which we impose
shortly before the axion mass becomes important (we comment at the end
of this subsection on our choice of the initial temperature). The axion
field $\theta(\vec x)$ takes on random values in different Hubble
patches. Therefore, we consider it as a random angular field with a
flat probability distribution function $f(\theta) = 1/(2\pi)$ for
$\theta \in [-\pi,\pi]$ and zero otherwise.\footnote{There is a
  subtlety related to this choice for $f(\theta)$, since being a
  random angular variable, any constant interval of length $2\pi$
  should be equivalent. Our choice is motivated by the fact that we
  adopt the harmonic approximation for the potential, which no longer
  is periodic. It turns out that in this case the flat distribution in
  the symmetric interval $[-\pi,\pi]$ is the only physically
  meaningful choice. Any other interval would lead to unphysical
  implications of the zero mode.} As usual, the expectation value of
any quantity $Y(\theta)$ is given by $\langle Y \rangle = \int d\theta
\, f(\theta) Y(\theta)$.  In particular, it implies for the mean and
the variance:
\begin{equation}\label{eq:expect}
\langle\theta(\vec{x})\rangle = 0 \,,\qquad \langle \theta(\vec{x})^2\rangle =
\pi^2/3 \,.  
\end{equation}

Let us now consider the Fourier transform
\begin{equation}\label{eq:fourier}
  \theta_k =  \int_V d^3x \, \theta(\vec{x}) e^{i\vec{k}\vec{x}} \,,\qquad
  \theta(\vec{x}) = \frac{1}{(2\pi)^3} \int d^3k \, \theta_k e^{-i\vec{k}\vec{x}} \,.
\end{equation}
The integral over $d^3x$ is taken over a large volume $V$, such that
the integral is finite, and $\vec x$ and $\vec k$ are co-moving
coordinate and momentum, respectively. We have $\langle \theta_k
\rangle = 0$, and $\theta_{-k} = \theta_k^*$ since $\theta(\vec{x})$
is real. Due to statistical homogeneity and isotropy the correlation
function in Fourier space can be written as
\begin{align}
  \langle \theta_k \theta^{*}_{k'} \rangle &=
  (2\pi)^3 \, \delta^3(\vec k - \vec k') P_\theta(k)
  \,,\label{eq:auto-k}
\end{align}
where $P_\theta(k)$ denotes the power spectrum for the field, which is
the Fourier transform of the 2-point correlation function $\xi(|\vec r|) =
\langle \theta(\vec x)\theta(\vec x + \vec r) \rangle$. We
follow the conventions for the power spectrum of
\rcite{Zentner:2006vw}.

We can now use the shape of the power spectrum to implement that causally disconnected regions should be uncorrelated. Let us introduce a characteristic wave number
\begin{equation}\label{eq:K}
  K = a_i H_i \,,
\end{equation}
where $a_i$ is the scale factor at our initial time $t_i$ and $H_i$ is
the Hubble rate at that time. The axion field should be uncorrelated
at co-moving distances larger than $1/K$. Note that there is an
ambiguity in this definition. Alternatively we could use the
association of wave number and co-moving distance as $k = \pi/R$, which would
lead to an additional factor $\pi$ in \cref{eq:K} for $R =
1/(a_iH_i)$. In general, $K$ is defined only up to factors of order one,
which unfortunately introduces a large uncertainty, since $K$ enters
in many quantities of interest with third power.
  
The normalization of the power spectrum is fixed by requiring $\langle
\theta(\vec{x})^2\rangle = \pi^2/3$ according to \cref{eq:expect}. The
shape of the power spectrum should be determined by the evolution of
the field from the PQ scale down to the QCD scale. In absence of a
full simulation over so many orders of magnitude, we are forced to
make some (physically motivated) guesses. A reasonable assumption
seems to be a white noise (i.e., flat) power spectrum with a sharp
cut-off at co-moving wave number $K$ (``top-hat''):
\begin{equation}\label{eq:P-TH}
  P_\theta^{\rm TH}(k) = \frac{2\pi^4}{K^3}\Theta(K - k) \,.
\end{equation}
This means that fluctuations for each mode up to $K$ are equally
likely. However, the finite cut-off leads to an oscillating two-point
correlation function $\xi(r)$ which decreases only with the inverse of the
distance-squared, and hence, implies long-range correlations in configuration
space beyond the horizon. Therefore we consider as alternative a Gaussian
suppression of high wave numbers: 
\begin{equation}\label{eq:P-G}
  P_\theta^{\rm G}(k) = \frac{8\pi^4}{3\sqrt{\pi} K^3} \exp\left(-\frac{k^2}{K^2}\right) \,,
\end{equation}
which leads to exponential suppression of correlations also in
configuration space. Therefore, the Gaussian power spectrum seems to
be physically better motivated and we adopt it as our default
assumption. We will, however, also study the $k$-space top-hat power
spectrum, since it provides a sharp cut-off to all the integrals in
the following, making the effect of the scale $K$ more transparent.

\Cref{eq:auto-k} together with our assumptions on the power spectrum,
\cref{eq:P-G} respectively \cref{eq:P-TH}, serve as initial condition for the
field evolution which we consider in the following. Before proceeding
let us comment on the choice of our initial time $t_i$, or the
corresponding temperature $T_i$. We want to set $T_i$ above the scale
when the axion mass becomes important, in order to capture this
process correctly by solving the equation of motion. On the other
hand, we cannot set $T_i$ much higher, since our formalism does not
describe the effect of the topological strings, which are essential
for describing the random massless field. Therefore, we chose to set
$T_i = 3 T_{\rm osc}$, with $T_{\rm osc}$ determined by
\cref{eq:Tosc}. The actual value depends on the chosen axion mass, but
typical values are $T_{\rm osc} \simeq 1$~GeV. Since this energy scale
appears profusely in our calculations, we will present our results in
units of the wavenumber $K_1=a_1 H_1$ or the co-moving distance $R_1 =
K_1^{-1}$, where $a_1$ and $H_1$ are evaluated at the temperature of
1~GeV.

\subsection{The average energy density}

Sticking to the quadratic potential, the energy density of the axion field is given by
\begin{align}
  \rho(\vec x) = \frac{\fPQ^2}{2}\left[\dot\theta^2 - \frac{1}{a^2}(\vec\nabla\theta)^2 + m^2\left(T\right)\theta^2\right] \,.
\end{align}
Since the evolution equation is linear in the harmonic approximation,
the Fourier modes evolve independent according to \cref{eq:eom-k} and we can write
\begin{align}\label{eq:initial_cond}
  \theta_k(a) = \theta_k \, f_k(a) \,.
\end{align}
Here $\theta_k \equiv \theta_k(a_i)$ denotes the initial condition for
the field at the time $t_i$ and $f_k(a)$ is a real function encoding
the time (or $a$) dependence obtained from solving the equation of
motion with the initial condition $f_k(a_i) = 1$. The random
properties of the field characterized by Eq.~\eqref{eq:auto-k} are
thus encoded in the initial conditions $\theta_k$. We solve
\cref{eq:eom-k} numerically for a large set of modes, for details see
\cref{app:EOM}. We use the susceptibility $\chi(T)$ as well as the
effective number of degrees of freedom as a function of temperature
needed to determine $H(T)$ from the QCD calculations from
\rcite{Borsanyi:2016ksw}, see also \rcite{Kitano:2015fla,
  diCortona:2015ldu} for similar calculations.

With this notation we obtain for the energy density
\begin{align}\label{eq:rho}
  \rho(\vec x) = \frac{1}{(2\pi)^6} \frac{f^2_{\rm PQ}}{2}  \int
  d^3k d^3k'\, \theta_k \theta^{*}_{k'} F(k, k') e^{-i\vec
    x(\vec k - \vec k')} \,,
\end{align}
where we have defined
\begin{align}\label{eq:Fkkp}
  F(k,k') = \dot f_k \dot f_{k'} + \left( \frac{\vec k\cdot\vec k'} {a^2} +  m^2\left(T\right) \right)f_k f_{k'} \,.
\end{align}
The average energy density is obtained by using the correlator from \cref{eq:auto-k} as
\begin{align} \label{eq:rho-bar}
  \overline\rho \equiv \langle\rho(\vec x)\rangle =
   \frac{1}{2\pi^2} \frac{f^2_{\rm PQ}}{2} \int_0^\infty dk \, k^2 \, P_\theta(k) F(k, k) \,,
\end{align}
with
\begin{align}
  F(k,k) = \dot f^2_k + \omega_k^2 f_k^2 \,,
\end{align}
where $\omega_k$ has been defined in \cref{eq:eom-k} and it
can be identified with the energy of the mode with
momentum $\vec k$.  Since the power spectrum suppresses
modes with $k > K$, at sufficiently late times, the term $k^2/a^2$ can
be neglected compared to the zero-temperature mass $m$. We say that all
modes become non-relativistic.

\begin{figure}
\centering
\includegraphics[width=0.85\textwidth]{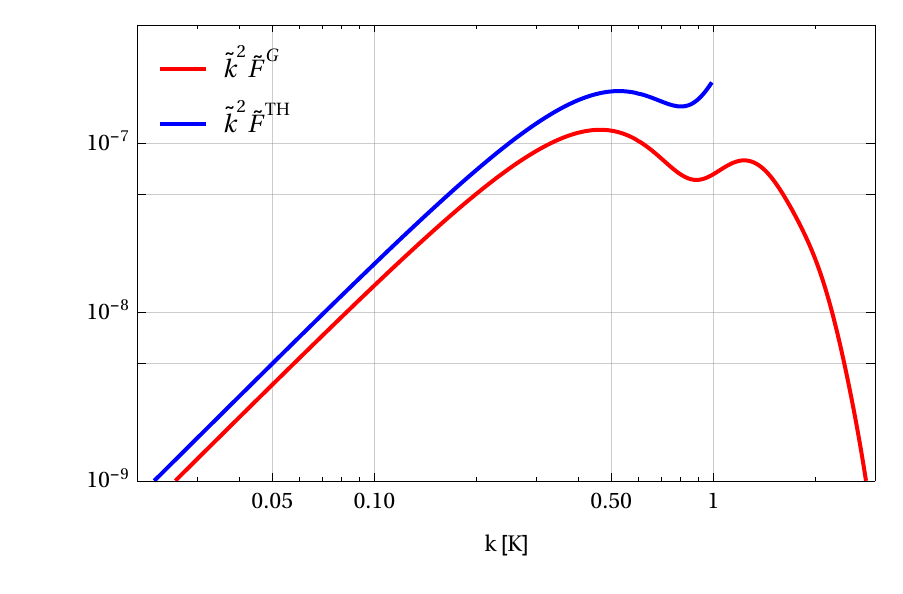}  
  \caption{Contributions to the average energy density according to
    Eq.~\eqref{eq:rho-bar2} when using the top-hat (TH) (blue) respectively the Gaussian (G) (red)
    power spectrum for the axion field. For the plot we chose $\fPQ=10^{12}\,{\rm GeV}$. 
    \label{fig:Fkk}}
\end{figure}

Let us introduce the dimensionless wave number
$\tilde k = k/K$.  
It follows from the equation of motion that once all relevant modes
have become non-relativistic and $m(T)$ reached its zero-temperature
value, $F$, and consequently $\overline\rho$, scales as $a^{-3}$, as
it should for cold dark matter. We factor out the $a^{-3}$ dependence
and use $m_0^2$ in order to define a dimensionless quantity
$\tilde{F}$ through $F = m_0^2 (a_\star/a)^3 \tilde F$, with $a_\star$
corresponding to $T_\star = 100$~MeV. Assuming for illustration the
top-hat power spectrum defined in \cref{eq:P-TH}, we find
\begin{equation}\label{eq:rho-bar2}
  \overline\rho = 
   \frac{\fPQ^2}{2} m_0^2 \left(\frac{a_\star}{a}\right)^3 
     \pi^2 \int_0^1 d\tilde k \, \tilde k^2 \, \tilde F( \tilde k, \tilde k) 
   \quad\qquad(P^{\rm TH}_\theta)\,. 
\end{equation}
We illustrate the contribution of the $k$-modes in \cref{fig:Fkk} for
both the top-hat as well as the Gaussian correlator. While the former
just cuts off modes with $k > K$, the latter provides a smooth
suppression. Note that the combination $\fPQ^2m_0^2$ in
\cref{eq:rho-bar2} is independent of $\fPQ$; the $\fPQ$ dependence is
hidden in this expression in the function $F(k,k)$, whose shape and
normalization depends on $T_{\rm osc}$ which in turn depends on
$\fPQ$. \Cref{eq:rho-bar2} agrees parametrically with the classical
result,
e.g.~\rcite{Preskill:1982cy,Abbott:1982af,Dine:1982ah,Turner:1985si}:
\begin{align} \label{eq:rho-pre-infl}
  \rho \sim \fPQ^2 m(a_{\rm osc}) m_0 \left(\frac{a_{\rm osc}}{a}\right)^3 \theta_{ini}^2\,,
\end{align}
where $\theta_{ini}$ denotes some ``initial'' mis-alignment angle. The
usual assumption in the post-inflation scenario, $\theta_{ini}^2 =
\pi^2/3$, is replaced in our result by the proper weighted
contribution of non-zero $k$-modes according to the initial power
spectrum.

Numerically we find for the current energy density of the axions
relative to the critical density due to the re-alignment mechanism in
the stated approximations:
\begin{equation}
  \Omega h^2
  \approx 0.1 \left(\frac{\fPQ}{10^{12}\,{\rm GeV}}\right)^{7/6}   
  \approx 0.1 \left(\frac{m_0}{5.7\times 10^{-6}\,{\rm eV}}\right)^{-7/6} \,.  
\end{equation}
This is in good agreement with other recent results, for instance
\rcite{Bae:2008ue,Visinelli:2009zm,Wantz:2009it}.  The numerical
coefficient 0.1 depends somewhat on our assumptions.  The energy
density is about a factor 2 larger when we change the initial
temperatur from $T_i = 3T_{\rm osc}$ to $T_i = 2T_{\rm osc}$ because
$k$-modes have less time to red-shift before they become
non-relativistic. The dependence on the chosen field correlator (Gauss
versus top-hat) is less than 30\%. In the following we will show
results for $f_\PQ$ equal to $10^{10}$, $10^{11}$ and $10^{12}$~GeV,
spanning approximately the range where the re-alignment mechanism
provides 0.5\% to 100\% of the dark matter energy density.

\subsection{Axion energy density power spectrum}

In this subsection we will compute the power spectrum of the density contrast of the axion field. For the Fourier transform of the density we find
\begin{align}
  \rho_q &=  \frac{1}{(2\pi)^3} \frac{f^2_{\rm PQ}}{2} \int d^3k \, \theta_k \theta^*_{k-q}
  F(k, k-q) \,,
\end{align}
with $\langle\rho_q\rangle \propto \delta^3(\vec q)$ following from \cref{eq:auto-k}.
Now we consider the density contrast:
\begin{align}
  \delta(\vec x) \equiv \frac{\rho(\vec x) - \overline \rho}{\overline \rho} \,,
\end{align}
with its Fourier transform $\delta_q = \rho_q / \overline\rho$ ($q
\neq 0$). The power spectrum is related to the variance of $\delta_q$
by $P(q) = \langle |\delta_q|^2\rangle / V = \langle
|\rho_q|^2\rangle/(V \overline\rho^2)$, see e.g.,
\rcite{Zentner:2006vw}.  Hence, we calculate:
\begin{align}
  \langle |\rho_q|^2\rangle = 
  \left[ \frac{1}{(2\pi)^3}\frac{f^2_{\rm PQ}}{2}\right]^2
  \int d^3k d^3k'\, \langle\theta_k \theta^*_{k-q}\theta_{k'}^* \theta_{k'-q}\rangle
  F(k, k-q)F^*(k', k'-q) \,.
\end{align}
With Wick's Theorem one obtains
\begin{align}
  \langle\theta_k \theta^*_{k-q}\theta_{k'}^* \theta_{k'-q}\rangle &=
  \langle\theta_k \theta^*_{k-q} \rangle\langle \theta_{k'}^* \theta_{k'-q}\rangle +
  \langle\theta_k \theta_{k'}^* \rangle\langle \theta^*_{k-q}\theta_{k'-q}\rangle +
  \langle\theta_k \theta_{k'-q} \rangle\langle\theta^*_{k-q}\theta_{k'}^* \rangle \\
  &= (2\pi)^6 P_\theta(|\vec k|) P_\theta(|\vec k -  \vec q|)
    \left\{ [\delta^3(\vec k-\vec k')]^2 +
            [\delta^3(\vec k+\vec k' - \vec q)]^2\right\} \,.
            \label{eq:wick}
\end{align}
where we have used eq.~\eqref{eq:auto-k} and we have droped terms with
$\delta^3(\vec q)$ by assuming $q \neq 0$. In
order to deal with the squares of the Dirac delta function we use
$\delta^3(k=0) = V/(2\pi)^3$. The first term in the curle bracket of \cref{eq:wick}
gives $|F(k,k-q)|^2$.
For the second term we can use that $F(q-k,-k) = F(k,k-q)$, which
follows from $f_k = f_{-k}^*$. Hence the first and second terms are
equal  and we obtain
\begin{align}
  \langle |\rho_q|^2\rangle = 
  2  \frac{ V}{(2\pi)^3}  \left(\frac{f^2_{\rm PQ}}{2} \right)^2
  \int d^3k \, P_\theta(|\vec k|) P_\theta(|\vec k -  \vec q|) \, F(k,k-q)^2  
\end{align}
Using eq.~\eqref{eq:rho-bar} for $\overline\rho$
this gives for the power spectrum
\begin{align}\label{eq:P}
  P(q) = \frac{1}{V}\frac{\langle |\rho_q|^2\rangle}{\overline \rho^2} =
  2 (2\pi)^3 \frac{\int d^3k \, P_\theta(|\vec k|) P_\theta(|\vec k -  \vec q|) \, F(k,k-q)^2}
  {\left[ \int d^3k \, P_\theta(k) F(k, k)\right]^2 }\,.
\end{align}

\begin{figure}
\centering
\subfigure{\includegraphics[width=0.49\textwidth]{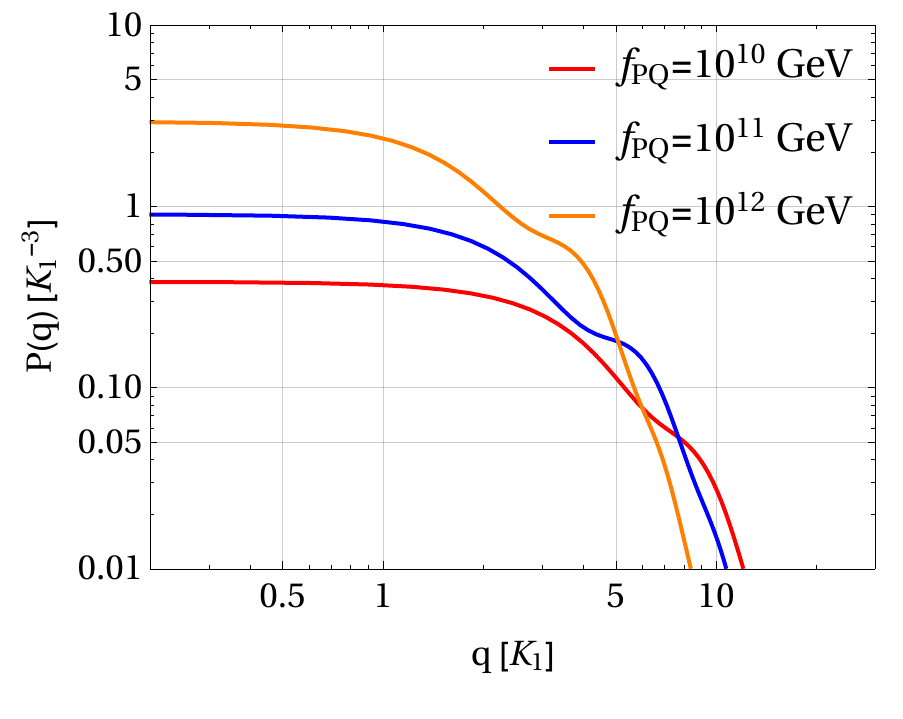}}
\subfigure{\includegraphics[width=0.49\textwidth]{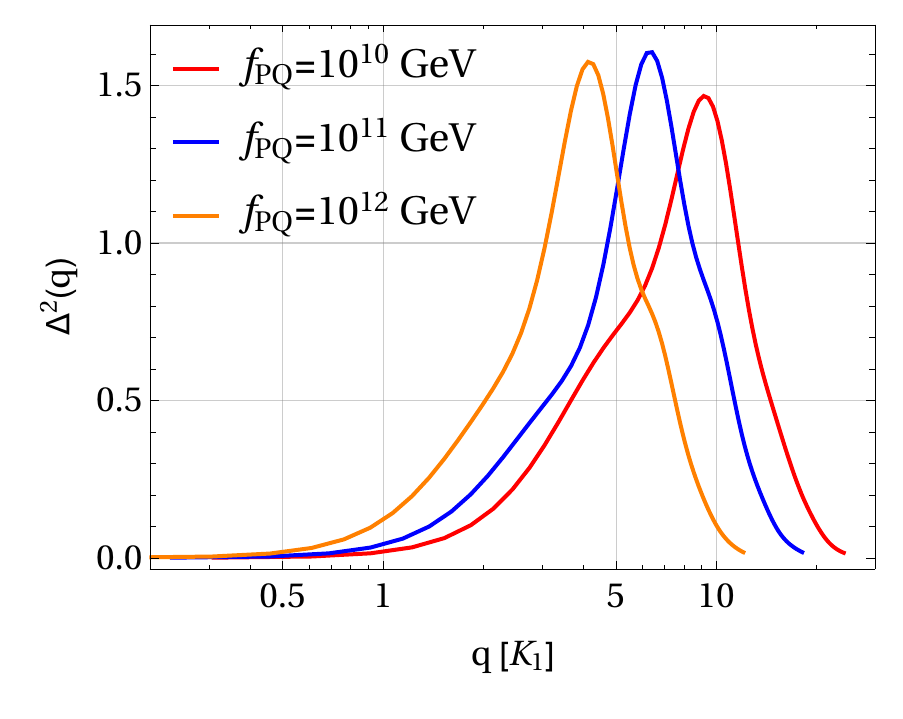}} 
\caption{The axion energy density power spectrum $P(q)$ (left) and the
  dimensionless power spectrum $\Delta^2(q)$ defined in
  \cref{eq:Delta} (right), for different choices of $\fPQ$, assuming the Gaussian initial axion field correlator. As reference scale we use the comoving wave number at 1~GeV: $K_1 = a_1H_1$. 
  \label{fig:Delta}}
\end{figure}

The function $F(k, k')$ defined in \cref{eq:Fkkp} is obtained from
solving the equation of motion as described in \cref{app:EOM} and it
depends on time. Once the axion has reached its zero-temperature mass
and all relativistic modes have been red-shifted away, $F(k, k')$
scales as $a^{-3}$, independent of $k, k'$. Hence the time dependence
in numerator and denumerator of \cref{eq:P} cancels and the power
spectrum becomes constant in time. Our numerical calculation shows that for
temperatures below
\begin{equation}
  T_\star \equiv 100 \, \rm MeV
\end{equation}
this is indeed the case. Since for the following considerations we
only need the power spectrum, we can stop the field evolution at that
point.  In the left panel of \cref{fig:Delta} we plot the power
spectrum at $T_\star$ for different choices of $\fPQ$. We observe constant
power at small $q$ (large scales), corresponding to white noise. Then
the power drops at a characteristic scale, corresponding roughly to the size
of the miniclusters, and is suppressed for large wave numbers (small
scales) where fluctuations are erased by the gradient terms.

In the right panel of \cref{fig:Delta} we show the dimensionless power
spectrum
\begin{equation}\label{eq:Delta}
  \Delta^2(q)=\frac{q^3}{2\pi^2} P(q) \,,
\end{equation}
which corresponds to the variance of the relative density
perturbations per decade of $q$. From the plot we see that relative
density fluctuations are of order one, i.e., non-linear.  Furthermore,
the peak in the relative density fluctuations is at a characteristic
wave number corresponding to a scale a few times smaller than the
horizon at $T_{\rm osc}$. This can be seen for instance by considering
the orange curve, corresponding to $\fPQ = 10^{12}$~GeV. For this
case, $T_{\rm osc} \approx 1$~GeV, and hence $K_1 = a_1H_1$ is the
inverse of the horizon at $T_{\rm osc}$. The peak for the orange curve
is around $q\approx 4K_1$, and hence it corresponds to a size 4 times
smaller than the horizon at $T_{\rm osc}$. For the other two curves,
$\fPQ$ is smaller, which means larger $T_{\rm osc}$, and therefore the
peak is shifted to smaller length scales accordingly.

\begin{figure}
\centering
\subfigure{\includegraphics[width=0.49\textwidth]{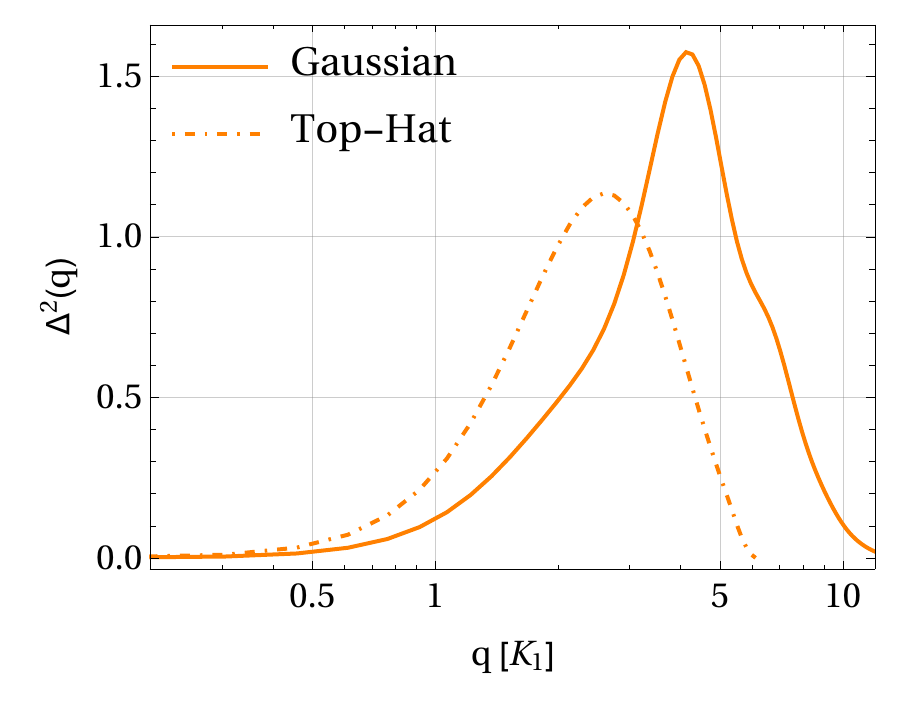}}
\subfigure{\includegraphics[width=0.49\textwidth]{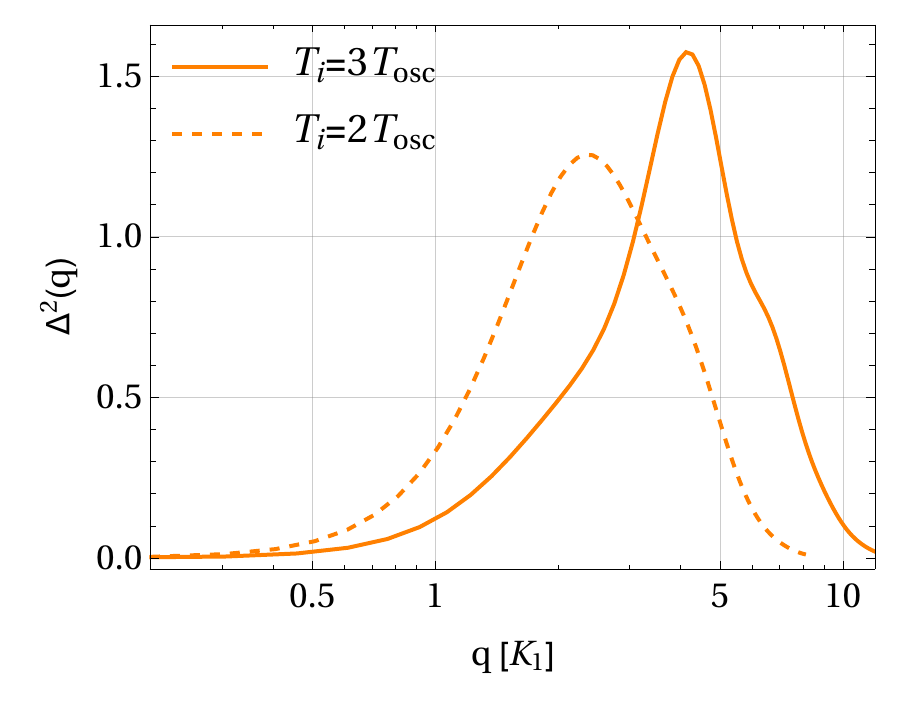}} 
\caption{The
  dimensionless power spectrum $\Delta^2(q)$ for $\fPQ=10^{12}$~GeV. In the left panel we compare $\Delta^2(q)$ for the different initial field correlators, Gaussian (G) and top-hat (TH), as defined in \cref{eq:P-G} respectively \cref{eq:P-TH}. The right panel shows $\Delta^2(q)$ assuming different inital times $T_i$, when using the same correlator (G). The reference scale, $K_1=a_1H_1$, is the comoving wave number at 1~GeV. The solid curve in both panels corresponds to our default assumption, $T_i=3T_{\rm{osc}}$ and Gaussian correlator, and is the same as the orange solid curve in the right panel of \cref{fig:Delta}. 
  \label{fig:Delta2}}
\end{figure}

In \cref{fig:Delta2} we show the impact of our initial assumptions on $\Delta^2(q)$. An interesting result is that the power spectrum has a cut-off around
$2K$ (instead of the naively expected $K$). This is most transparent
for the case when we consider a top-hat initial correlator for the
axion field according to \cref{eq:P-TH}, where we have a sharp cut-off
in $k$-space. For $\fPQ=10^{12}$~GeV we have $K/K_1 = a_i H_i/(a_1 H_1) \approx
a_1/a_i \approx 3$, since $T_i/(1~{\rm GeV}) \approx T_i/T_{\rm osc} = 3$.
Therefore, the value $q/K_1\approx 6$, at which the
dash-dotted curve in the left panel of \cref{fig:Delta2} goes to zero corresponds to $2K$. This result
follows directly from the way how the two $P_\theta$ factors in
\cref{eq:P} depend on the wave number, and it implies that although modes
with $k>K$ do not contribute to the energy density, there is power in
fluctuations up to wave numbers $2K$. Note that for the Gaussian
correlator, \cref{eq:P-G}, which is our default assumption, the
cut-off is smeared out.

The comparison of the solid and dashed curves in the right panel of \cref{fig:Delta2}
shows the effect of changing our default assumption $T_i = 3T_{\rm
  osc}$ to $T_i = 2T_{\rm osc}$. Note that this implies also a change
of the wave number cut-off, which we define as $K=a_iH_i$. As expected
we observe a shift of the peak towards smaller wave numbers.

A note on the normalization of our power spectrum is in order. We
  use $\overline\rho$ to normalize the spectrum, which is the average
  density from the re-alignment mechanism. If there is an additional
  contribution to the axion energy density (e.g., from the string and
  domain wall decay) the power would be reduced accordingly, unless
  the additional component itself introduces further fluctuations.

\bigskip

Our calculations so-far do not include the effect of gravity on the
axion over-densities, therefore the expression for the power spectrum,
\cref{eq:P} remains constant after $T_\star$. In the following we are
going to ``switch on'' gravity for the axions, and develop a model to
describe the over-densities up to the point when they decouple from the
Hubble expansion. The power spectrum at $T_\star$ discussed in this
section will be the input for those calculations.

\section{The size and mass of axion miniclusters}
\label{sec:distr}

In standard cold dark matter cosmology, the
Press \& Schechter \cite{Press:1973iz} (PS) method and its
variants are useful tools to estimate the
mass function of gravitationally collapsed objects (for reviews see e.g., \rcite{Zentner:2006vw,
  Maggiore:2009rv}). The basic idea is to use a spherical collapse
model for an over-density to estimate a critical density contrast,
$\delta_c$, such that regions with $\delta > \delta_c$ are collapsed. PS provide a rule to use this result to estimate the mass function, $dn/dM$,
which is the number density of collapsed objects with mass in the
interval $[M, M+dM]$. Below we provide a modification of the standard
method to take into account several peculiarities of the small-scale
fluctuations in the axion energy density:
\begin{enumerate}
\item As is clear from the previous section, density fluctuations are of
  order one from the initial moment when they are created. Therefore,
  linear theory cannot be used.
\item The non-linear fluctuations are created around $T\sim 1$~GeV,
  deep inside the radiation dominated era. We need a collapse model
  which is valid both during radiation and matter domination.
\item We will be interested in the double differential mass function
  $dn/dMdR$, providing the density of objects with a certain mass and a
  certain size. We develop a modified PS approach to calculate
  $dn/dMdR$, taking again into account the non-linearity of the
  fluctuations.
\end{enumerate}

In \cref{sec:collapse,sec:MF} we present the spherical collapse model
and our derivation of the double differential mass function under
those requirements, with the main result for $dn/dMdR$ given in
\cref{eq:MF}. The reader mostly interested in the application of that
result to axions may directly skip to \cref{sec:MCresults}, where we
present the results of our calculation for the minicluster
distribution around matter-radiation equality.

\subsection{Spherical collapse model}
\label{sec:collapse}

Kolb \& Tkachev (KT) \cite{Kolb:1994fi} provide a method to describe
spherical collapse during and after radiation domination. Here we do
not repeat their calculation but just present briefly the approach and
state the results, which we are going to apply in the following.  The
equation of motion for a spherical shell of matter, including a homogeneous radiation background energy density, $\rho_{\rm rad}$, is described by the differential equation
\begin{equation}
\ddot{r}=-\frac{8\pi G}{3}\rho_{\rm rad} r - \frac{GM}{r^2}\, ,
\end{equation}
where $r$ is the physical radius and the total dark matter mass $M$
enclosed in the sphere of radius $r$ is assumed to remain constant
during collapse. Let us denote by $r_{\rm flow}$ the physical
coordinate describing the background expansion. Then we introduce the
dimensionless variable $\xi$ to describe the deviation of the
over-density from this expansion: $r=\xi r_{\rm flow}$. KT derived an
equation of motion for $\xi$:
\begin{equation}\label{eq:xievo}
x(1+x)\frac{d^2\xi}{dx^2} + \left(1+\frac{3}{2}x\right)\frac{d\xi}{dx}+\frac{1}{2}\left(\frac{1+\delta}{\xi^2}-\xi\right)=0\, ,
\end{equation}
where $x\equiv a/a_{\rm eq}$, with $a_{\rm eq}$ being the scale factor at matter-radiation equality.
The density contrast $\delta$ is the over-density at the initial time where we start the evolution. It is related to $M$ through
\begin{equation}\label{eq:Mr}
M = \frac{4\pi}{3}\overline{\rho}\left(1+\delta\right)r^3 \,,
\end{equation}
with $r$ denoting the initial size of the over-dense region.
Eq.~\eqref{eq:xievo} is valid both in the radiation and matter
domination era. The solution $\xi(x)$ of
\cref{eq:xievo} can be used to identify the time when an over-density
collapses by requiring $\dot r = 0$, i.e., when the over-density
``turns around'' and starts to contract. We have verified by
numerically solving \cref{eq:xievo} the result of KT, namely that an
initial over-density $\delta$ at an early time will turn around at $x$ if
$\delta > \delta_c$ with
\begin{equation}\label{eq:delta_c}
  \delta_c(x) \approx \frac{0.7}{x} \,.
\end{equation}
This result holds for $x<1$ (radiation domination) as well as $x>1$
(matter domination), and is to good approximation independent of the
initial time. As we have seen above, the minicluster power spectrum
remains constant shortly after all modes became non-relativistic and
the axion reaches its zero-temperature mass.  Hence, the precise point
when we start the spherical collapse is not important as long as the
corresponding temperature is less than $T_\star = 100$~MeV. For
definiteness, we set the initial time of the collapse calculation to
that temperature and denote initial quantities with the index $\star$.

\subsection{Double differential mass function}
\label{sec:MF}

\begin{figure}
\subfigure{\includegraphics[width=0.49\textwidth]{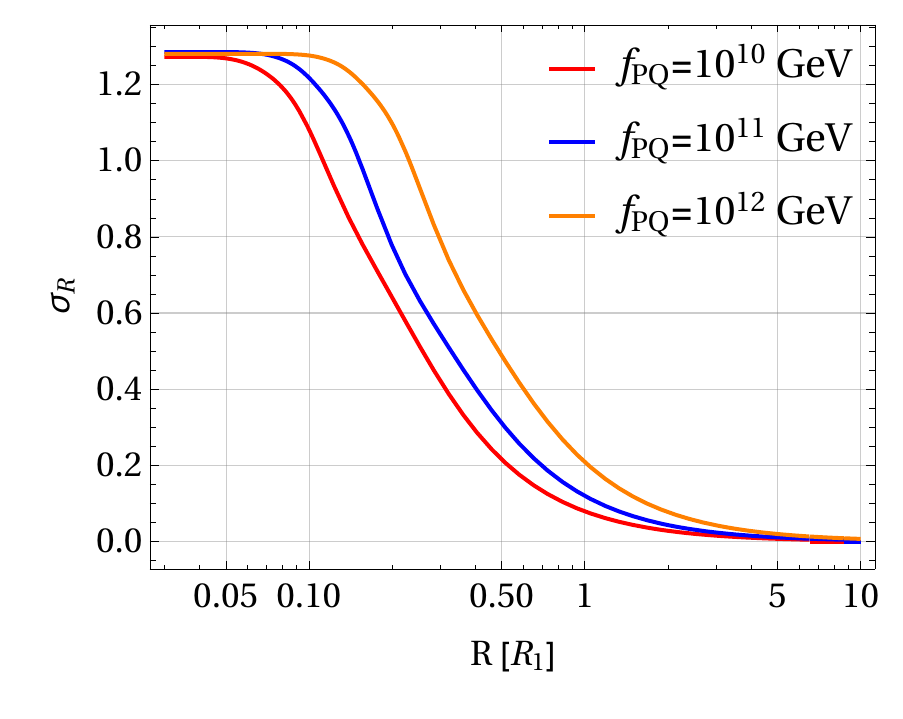}}
\subfigure{\includegraphics[width=0.49\textwidth]{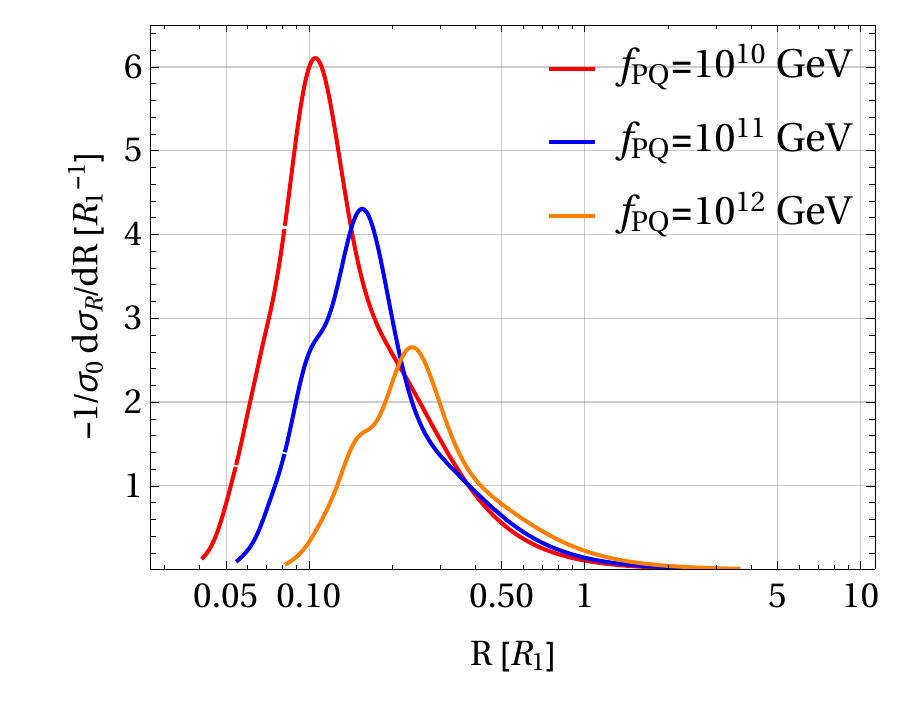}}
\caption{Standard deviation of the smoothed density field (left) and
  relative derivative of the standard deviation, $-
  1/\sigma_0d\sigma_R/dR$, (right) as a function of the smoothing scale $R$ for
  different choices of $\fPQ$. The reference length scale is the
  co-moving size of the horizon at 1~GeV: $R_1 = 1/(a_1H_1)$.}
  \label{fig:sigma_R}  
\end{figure}

Let us consider the axion energy density contrast smoothed over a characteristic length scale $R$:
\begin{equation}
  \delta_R(\vec{x}) = \int d^3x' \,W_R(\vec x - \vec x') \delta(\vec x') \,,
\end{equation}
where $W_R(\vec x)$ is a filter function which goes to zero if $x\gg R$. 
Then the variance of the smoothed density contrast is determined by the power spectrum:
\begin{equation}
\sigma_R^2 \equiv  \langle \delta_R(\vec x)^2 \rangle    
  = \frac{1}{2\pi^2} \int_0^{\infty} dk \,k^2 P(k) \left|\tilde W_R(k)\right|^2 \,,
\end{equation}
where $\tilde W_R(k)$ is the Fourier transform of the window
function. We adopt a top-hat window function in $k$ space: $\tilde
W_R(k) = \Theta(1 - kR)$. We comment on the reason for this choice
below. As visible in \cref{fig:sigma_R}, $\sigma_R$ has a step-like
shape with the characteristic scale ranging from 3 to 10 times smaller than the horizon at $T=1$~GeV, depending on the value of $\fPQ$. 

We are going to assume that $\delta_R(\vec{x})$ is a random Gaussian
variable with variance $\sigma^2_R$, i.e., the probability to find a fluctuation in the
smoothed energy density in the interval $[\delta,\delta + d\delta]$ is
\begin{equation}\label{eq:f_sm}
  f_\text{sm}(\delta;R) = \frac{1}{\sqrt{2\pi}\sigma_R}
  \exp\left(-\frac{\delta^2}{2\sigma_R^2}\right) \,.
\end{equation}
Since $\sigma_R$ is of order unity, large fluctuations are likely. The
Gaussian shape implies then, that the total density can become
negative. However, below we will be interested only in upward
fluctuations $\delta>0$, and therefore we are not applying
\cref{eq:f_sm} in the potentially unphysical region. Furthermore, the
Gaussian assumption for the fluctuations is consistent with using the
harmonic potential. Large over-densities due to anharmonic effects may also
lead to non-Gaussian tails of the distribution.

In the standard PS formalism a one-to-one correspondence between the
smoothing scale $R$ and the mass contained in the over-density is
assumed by the ansatz $M_R = V_R \overline\rho$, with $V_R$ being the
volume associated with the window function. Here we want to relax this
ansatz and allow for the fact that the variance of $\delta$ is
large. Therefore, the mass of an over-dense region depends both on the
size $r$ and the over-density $\delta$ via \cref{eq:Mr}.

Our goal is now, departing from \cref{eq:f_sm}, to derive the joint
probability distribution function (pdf) $f(\delta, r)$ for $\delta$ and $r$, which gives the probability to find a fluctuation with
$\delta \in [\delta, \delta + d\delta]$ which has a size $r\in[r,r+dr]$.
Note that \cref{eq:f_sm} is a pdf for $\delta$ at fixed $R$, normalized to 1 for any $R$.
The smoothing at scale $R$ implies that only fluctuations with
$r > R$ can contribute to the pdf of $\delta$. Hence we make the ansatz
\begin{equation}\label{eq:rel1}
  g(R) \, f_\text{sm}(\delta;R) = \int_R^\infty dr \, f(\delta,r) \,, 
\end{equation}
where the function $g(R)$ is introduced such that the marginal
distribution $f(r) \equiv \int d\delta\, f(\delta, r)$ is properly normalized:
\begin{align}
  g(R) = \int_R^\infty dr \, f(r) \,,\qquad g(0) = 1 \,.
\end{align}
In general this leads to complicated integro-differential equations for
the unknown functions $g(r)$ and $f(\delta, r)$.
However, using the Gaussian for $f_\text{sm}(\delta;R)$
from eq.~\eqref{eq:f_sm} we can try to guess the solution. 
By differentiating \cref{eq:rel1} one obtains
\begin{align}\label{eq:fDE}
  f_\text{sm}(\delta;R) \left[g'(R) - \frac{d \log \sigma_R}{d R}
    \left(1 - \frac{\delta^2}{\sigma_R^2}\right) g(R) \right] = - f(\delta, R)
\end{align}
Indeed, it is easy to show that $g(R) = \sigma_R/\sigma_0$
provides a solution, with $\sigma_0 \equiv \sigma_{R=0}$ being
the variance without smooting. Using \cref{eq:fDE} we obtain:
\begin{align}
  f(\delta,R) &= - \frac{1}{\sigma_0}\frac{d\sigma_R}{dR} \frac{\delta^2}{\sigma_R^2}
  f_\text{sm}(\delta;R) \,, \label{eq:f_del_r}\\
  f(R)&= - \frac{1}{\sigma_0}\frac{d\sigma_R}{dR}\,. \label{eq:f_r}
\end{align}
The result for the marginal distribution in \cref{eq:f_r} has an
intuitive interpretation: the distribution of the size of the
fluctuations is related to the change in the smoothing scale, and if
$\sigma_R$ is constant at a given scale $R$, there are no fluctuations
of size $r=R$ at that scale. We show some numerical examples of $f(R)$
for the axion miniclusters in \cref{fig:sigma_R}.

\bigskip

Combining our result for $f(\delta, r)$ with \cref{eq:delta_c}, we can
now proceed in analogy to the PS formalism and estimate the double differential mass
function. We use that for fixed $r$, \cref{eq:Mr} relates the mass $M$
to the over-density $\delta$. We denote by $dn/dMdR$ the comoving
number density of collapsed objects with mass in $[M,M+dM]$ and size
in $[R,R+dR]$. It is related to $f(\delta, r)$ by
\begin{equation}\label{eq:def-MF}
  \frac{M}{\overline\rho} \frac{dn}{dMdR} \, dM dR = 2 \, f(\delta, R)
  \, d\delta dR \, \Theta[\delta - \delta_c(x)] \,.
\end{equation}
The theta-function selects over-densities larger than
$\delta_c(x)$, which are collapsed at the time $x$. The factor of 2
is included here for the same reason as it appears in the original PS
formula. It takes into account the mass in under-dense regions; if all
mass was bound in collapsed objects (meaning $\delta_c = 0$) the
integral of the right-hand side of \cref{eq:def-MF} should give 1,
whereas without the factor 2 it would give only $1/2$. Using
\cref{eq:Mr} we obtain our final result for the double differential
mass function:
\begin{equation}\label{eq:MF}
  \frac{dn}{dMdR}  = \frac{3}{2\pi M R^3} \, f(\delta, R) \Theta[\delta - \delta_c(x)] \,,
\end{equation}
where $f(\delta, R)$ is given in \cref{eq:f_del_r}, $\delta$ is
considered as a function of $M$ and $R$, $\delta=\delta(M,R)$
according to \cref{eq:Mr}, and the critical density $\delta_c(x)$ is
given in \cref{eq:delta_c}. The interpretation of \cref{eq:MF} is as
follows: $dn/dMdR$ is the distribution of collapsed objects at a time
$x=a/a_{\rm eq}$, whereas $f(\delta, R)$ is the distribution of the
fluctuations at the initial time $x_\star$, which can be calculated
departing from the power spectrum at $x_\star$ using \cref{eq:f_del_r}.
The total mass function $dn/dM$ is obtained by integrating over $R$
\begin{equation}\label{eq:MF-int}
  \frac{dn}{dM}  = \frac{3}{2\pi M} \int_0^{R_c(M)} \frac{dR}{R^3} \, f[\delta(M,R), R] \, ,
\end{equation}
where $R_c$ for a given $M$ can be derived from \cref{eq:Mr} with $\delta = \delta_c(x)$.

Before applying this result to the axion minicluster, let us come back
to the question of how to chose the window function for smoothing the
energy density. Since the minicluster power spectrum has a
high-momentum cut-off, we do not expect to find structures at very
small scales. In the mass function this is reflected by the
proportionality to $d\sigma_R/dR$ via \cref{eq:f_del_r}. Indeed, from
\cref{fig:sigma_R} we observe that $d\sigma_R/dR$ goes to zero for
small $R$. However, from \cref{eq:MF} we see that $d\sigma_R/dR$ needs
to go sufficiently fast to zero for $R\to 0$ to compensate the factor
$1/(MR^3)$. It turns out that both for a Gaussian, as well as a
real-space top-hat window function, the mass function diverges for
small $M$ and $R$. Only the $k$-space top-hat indeed cuts off the
small structures, since $d\sigma_R/dR$ is exactly zero for $1/R$
larger than the cut-off in the power spectrum. This is a well known
problem also for the standard PS method, see, e.g.,
\rcite{Fairbairn:2017sil} for a recent discussion and further
references. Therefore, we use the $k$-space top-hat smoothing
function, which seems suitable to describe the physics of the power
spectrum cut-off.

\subsection{Minicluster mass and size distribution}
\label{sec:MCresults}

\begin{figure}[t]
\centering
\subfigure{\includegraphics[width=0.49\textwidth]{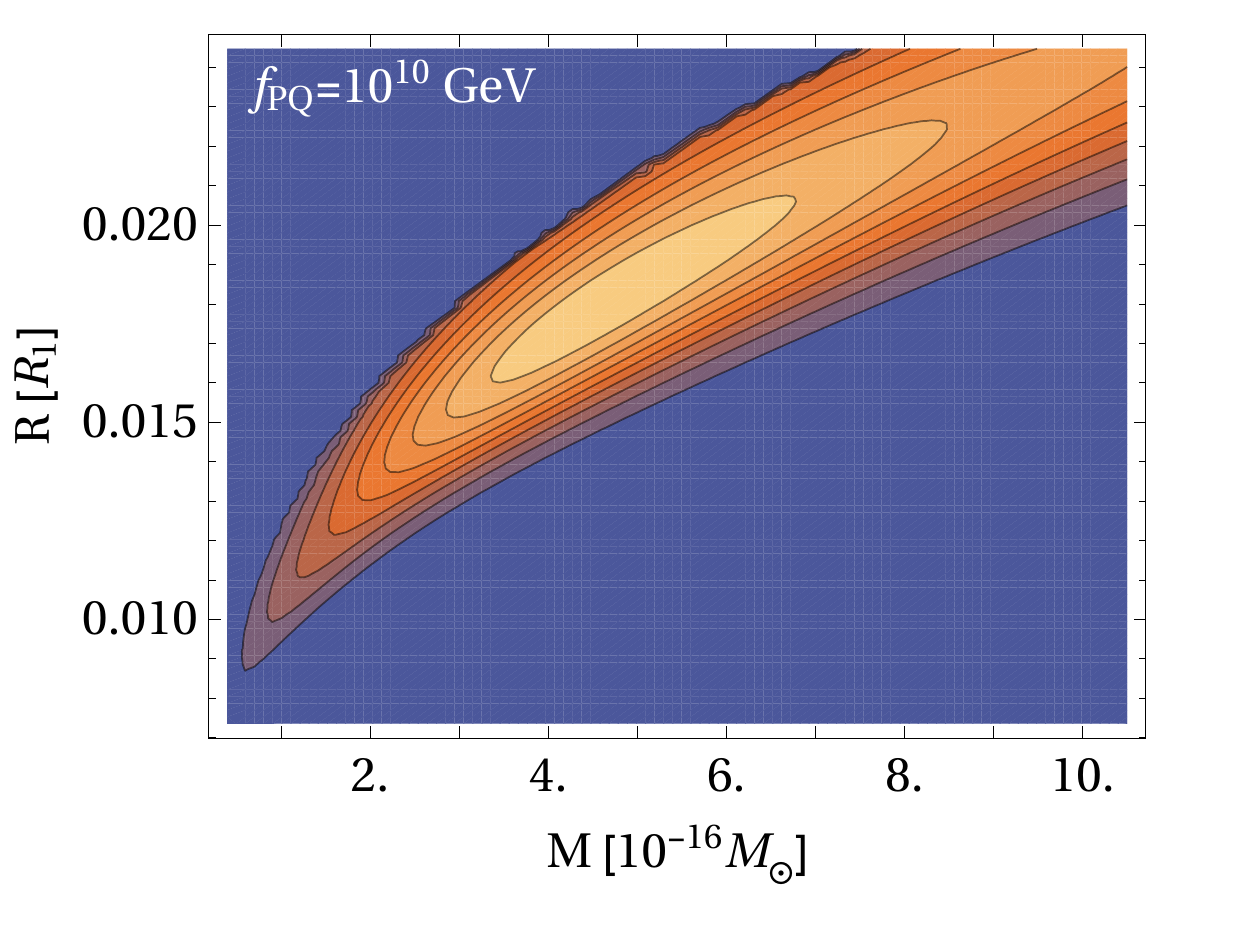}}
\subfigure{\includegraphics[width=0.49\textwidth]{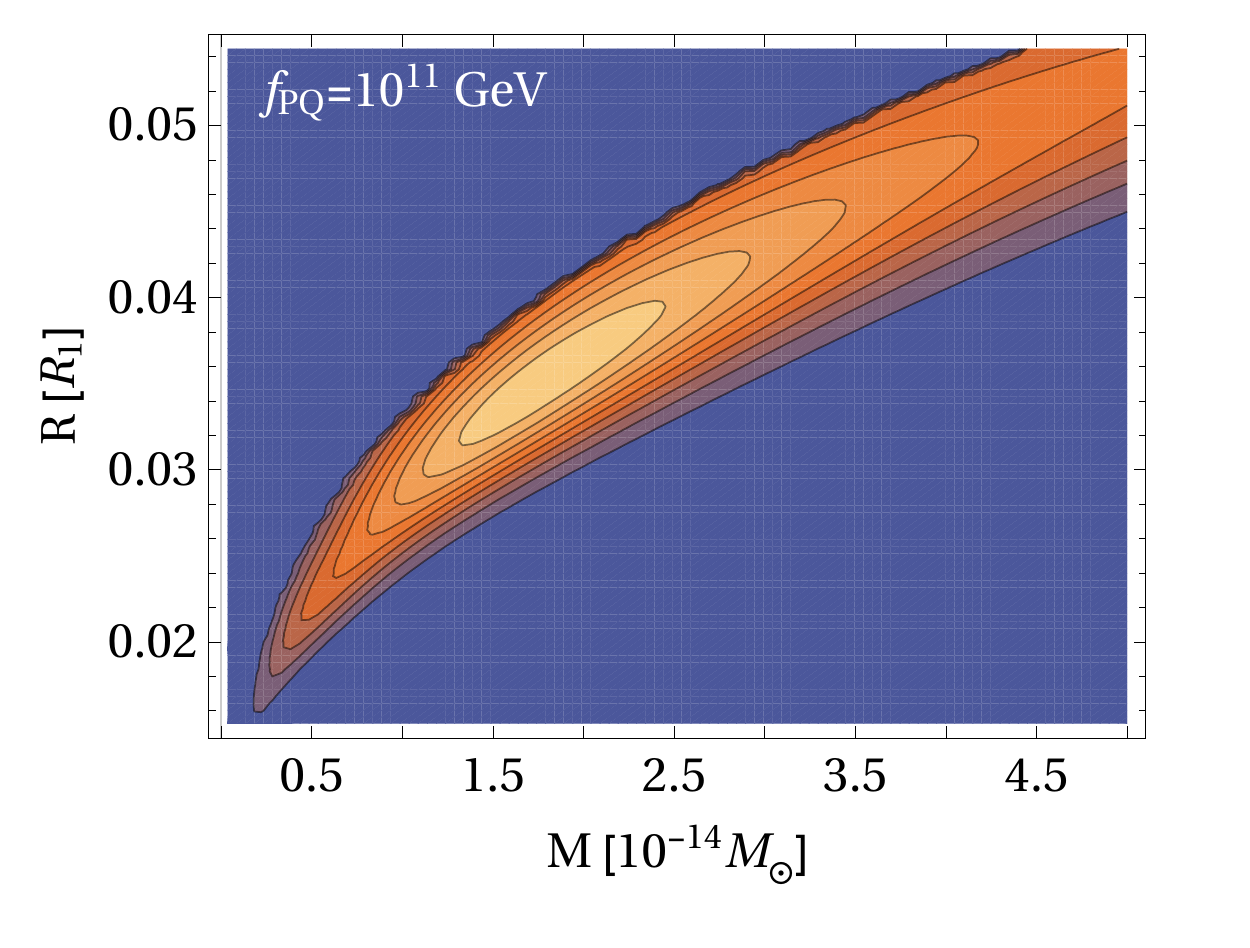}}
\subfigure{\includegraphics[width=0.59\textwidth]{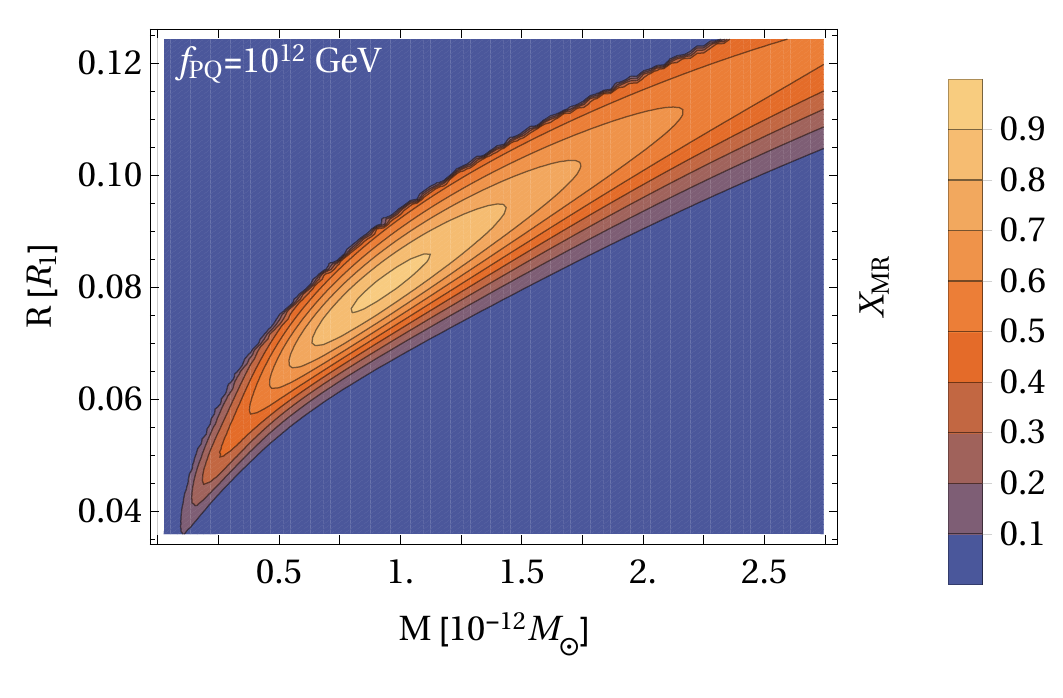}}
  \caption{Dimensionless double differential distribution of collapsed
    objecs $X_{MR}\equiv M^2R/\overline{\rho} (dn/dMdR)$ at
    matter-radiation equality for three choices of $\fPQ$. The
    vertical axis shows the co-moving size of the over-density at the
    initial time $T_\star = 100$~MeV relative to $R_1$, the co-moving Hubble
    radius at 1~GeV.}
  \label{fig:doublediff}  
\end{figure}

In order to display our results for the axion minicluster mass
function we introduce the dimensionless distributions
\begin{align}\label{eq:XMR}
X_{MR}=\frac{M}{\overline{\rho}} \, \frac{dn}{d \log M \, d \log R}
\end{align}
and
\begin{align}\label{eq:XM}
X_M=\frac{M}{\overline{\rho}}\,\frac{dn}{d \log M} \, .
\end{align}
They correspond to the contribution of objects per decade in $M$ and
$R$ for $X_{MR}$ and only in $M$ for $X_M$, relative to
$\overline\rho$. Note that $\overline\rho$ is the average energy
density from the re-alignment mechanism as obtained in
\cref{eq:rho-bar}, which in general will be smaller than the total
energy density in axions due to the string and domain wall decay
contribution. Hence, the possible presence of an additional
  energy density component will affect the normalization of the mass
  function as well as the power spectrum (and therefore
  $\sigma_R$). Below we will focus mostly on the shape of our
  distributions. In \cref{fig:doublediff} we show the double
differential distribution in $M$ and $R$ of collapsed miniclusters for
three choices of $\fPQ$. We observe a peaked distribution with a clear
correlation between mass and size of the objects.

\begin{figure}[t]
  \centering
  \includegraphics[width=0.85\textwidth]{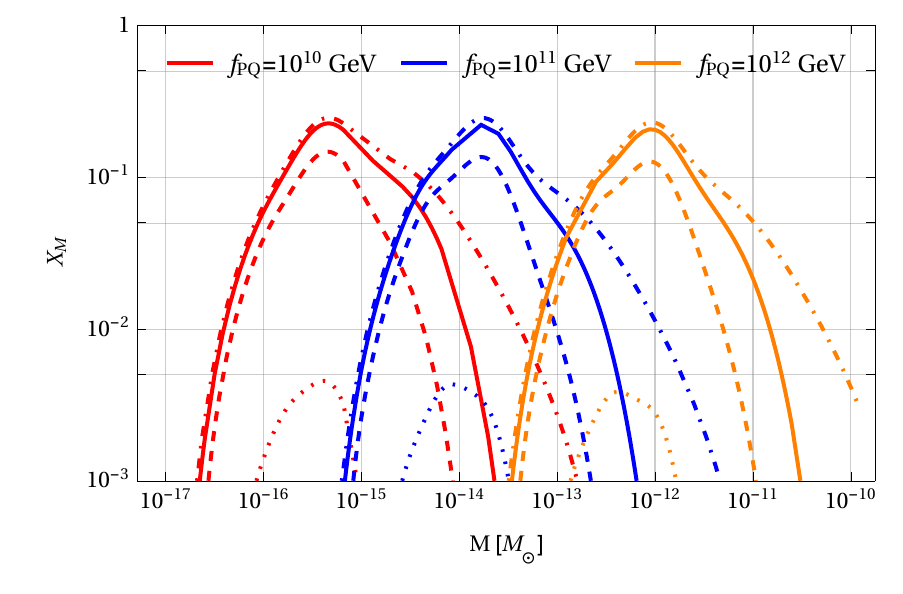}
  \caption{Dimensionless minicluster mass function $X_M \equiv
    M^2/\overline{\rho}(dn/dM)$ for three choices of $\fPQ$. The
    different line-styles indicate the mass function at different
    times: dotted $x=0.2$, dashed $x=0.5$, solid $x=1$, dot-dashed
    $x=5$, where $x = a/a_{\rm eq}$.}
\label{fig:masscontr}
\end{figure}

In \cref{fig:masscontr} we show the $R$-integrated mass function. At
matter-radiation equality (solid curves) we can make the following
observations: the location of the peak of the distributions depends
sensitively on $\fPQ$, ranging from $10^{-16}$ to $10^{-12} \,
M_\odot$. For given $\fPQ$, the range of masses which contribute more
than 1\% of the peak mass spans an interval of more than
two orders of magnitude in $M$ around the peak value, see
\cref{tab:peak} for the numbers.

\begin{table}
  \centering
  \begin{tabular}{ccccc}
    \hline\hline
    $\fPQ$ [GeV] & $M_{\rm peak} \, [M_\odot]$ & $M$ range $[M_\odot]$ & $r_{\rm ta}^{\rm peak}$ [km] & $r_{\rm ta}$ range [km]\\
    \hline
    $10^{10}$ & $4\times 10^{-16}$ &  $[2 \times 10^{-17},1 \times 10^{-14}]$ & $4\times 10^4$ & $[2\times 10^4, 2\times 10^5]$\\ 
    $10^{11}$ & $2\times 10^{-14}$ &  $[5 \times 10^{-16},3 \times 10^{-13}]$ & $2\times 10^5$ & $[4\times 10^4, 7\times 10^5]$\\
    $10^{12}$ & $8\times 10^{-13}$ &  $[6 \times 10^{-14},2 \times 10^{-11}]$ & $2\times 10^6$ & $[7\times 10^5, 7\times 10^6]$\\
    \hline\hline
  \end{tabular}
  \caption{For three example values of $\fPQ$ we give the minicluster
    mass for which the relative mass function $X_M$ peaks, $M_{\rm
      peak}$, and the interval in masses, where the mass function
    $X_M$ is larger than 1\% of the peak.  The column ``$r_{\rm
      ta}^{\rm peak}$'' gives the size of the over-density
    corresponding to $M_{\rm peak}$ when it decouples from the Hubble
    flow and starts to collapse (``turn-around''). The last column
    gives the range of $r_{\rm ta}$ corresponding to masses for which
    the mass function $X_M$ is larger than 1\% of the peak.}
   \label{tab:peak}
\end{table}

The different line-styles in \cref{fig:masscontr} show the mass
function at different times around matter-radiation equality, ranging
from $x=0.2$ till $x=5$. Note that with the normalization of the
distribution according to \cref{eq:XM} the expansion effect is
factored out and the plot shows the change of the number of objects per
co-moving volume. We find that the collapse process largely finishes
at matter-radiation equality ($x=1$, solid curves). For late times we
see some hierarchical collapsing at the high mass end. But we checked
that the dash-dotted curves ($x=5$) are already close to the
$x\rightarrow \infty$ limit.  This can be understood from the analytic
expression, \cref{eq:MF-int}, in the limit $\delta_c \to 0$.

Estimates of the minicluster mass in the previous literature assume
that a minicluster is made out of all axions inside the Hubble horizon
$d_H$ at the time the field oscillations commence \cite{Hogan:1988mp}:
$M \sim \frac{4\pi}{3} d_H^3(T_{\rm osc}) \overline\rho(T_{\rm
  osc})$. Using $d_H \sim 1/H$, this leads to (see e.g.,
\rcite{Tinyakov:2015cgg, Davidson:2016uok, Bai:2016wpg}) $M \sim 10^{-12} M_\odot
(\fPQ/10^{11}\,{\rm GeV})^2$. While our results show a similar
dependence on $\fPQ$, the values for $M_{\rm peak}$ obtained from
\cref{fig:masscontr} are about two orders of magnitude smaller. This
follows from the fact that the characteristic size of the density
fluctuations is smaller than the Hubble horizon at $T_{\rm osc}$.  Let
us consider the horizon in co-moving coordinates, $d_H/a = 1/(a H)$, at $T_{\rm osc}$
relative to our reference scale $R_1$:
\begin{equation}
  \frac{d_H(T_{\rm osc})}{a_{\rm osc}} \frac{1}{R_1} = \frac{a_1 H_1}{a_{\rm osc}H_{\rm osc}} =
  \left[0.49, 0.66, 1\right] \quad\text{for}\quad \fPQ =
  \left[10^{10},10^{11},10^{12}\right]~\rm{GeV} \,.  
\end{equation}
Considering \cref{fig:sigma_R,fig:doublediff}, those numbers
imply that the typical scale of the miniclusters is smaller than the
size of the horizon at $T_{\rm osc}$ and therefore we obtain lighter
miniclusters. Note that \rcite{Fairbairn:2017dmf} obtains an even
larger minicluster mass, since their definition of the ``Hubble
volume'' differs by a factor $\pi$ from the above estimate $d_H \sim
1/H$.

\bigskip

Let us now discuss the size of the miniclusters. The quantity shown on
the vertical axes of \cref{fig:doublediff} is not very intuitive: it
corresponds to the co-moving size of the over-density at the initial
time $T_\star = 100$~MeV relative to the co-moving Hubble radius at
1~GeV. In order to convert this into a more useful quantity, we
calculate now the physical size of an over-density of given mass, at
the time when it decouples from the Hubble flow, i.e., at turn-around, denoted by $r_{\rm ta}$.
In the notation of \cref{sec:collapse}, it is given by
\begin{equation}
r_{\rm ta} = \xi_{\rm ta} a_{\rm ta} R \,,
\end{equation}
where $R$ is the initial co-moving radius. By using \cref{eq:Mr}
and solving \cref{eq:xievo} numerically one can get $\xi_{\rm
  ta}$ and $a_{\rm ta}$. An approximate analytic expression can be
obtained by using \cite{Kolb:1994fi} $\xi \simeq 1- \delta x / 2$,
together with $\delta x_{\rm ta} \simeq 0.7$.  Introducing a minor
fudge factor to fit numerics we find
\begin{equation}
r_{\rm ta} \simeq 0.4 \frac{R a_{eq}}{\delta} 
= 0.4\frac{a_{eq}}{a_1 H_1} \, \frac{(R/R_1)^4}{M/M_1 - (R/R_1)^3} 
\label{eq:rta}
\end{equation}
where in the last step we use \cref{eq:Mr} to express $\delta$ in
terms of $R, M$ and a reference mass at $T_1=1$~GeV defined as $M_1
\equiv \frac{4\pi}{3} \overline\rho_1 H_1^{-3}$.  \Cref{eq:rta} makes
clear that a higher initial over-density leads to earlier collapse and
thus a smaller physical radius, compared to objects which are less
dense and therefore need a longer collapse time.  In the parameter
range relevant for axion miniclusters, we find that the $M/M_1$ term
in the denominator of \cref{eq:rta} is a factor of $10^2$ larger than
the $(R/R_1)^3$ term. Neglecting the latter then allows us to write
$r_{\rm ta}$ as
\begin{equation}
r_{\rm ta} \simeq 1.4 \times 10^9 ~{\rm km} \, \left(\frac{R}{R_1}
\right)^4\left(\frac{M_1}{M}\right) \,,
\end{equation}
where we have used $T_{\rm eq} = 0.8~\rm eV$.
$M_1$ depends on $f_\PQ$ via $\overline\rho$. For our three example values we obtain
\begin{equation}
  r_{\rm ta} \simeq \left[2, 25, 360\right] \times 10^{10} \, {\rm km}
  \left(\frac{R}{R_1} \right)^4\left(\frac{10^{-14} M_\odot}{M}\right) \,,\qquad \fPQ =
  \left[10^{10},10^{11},10^{12}\right]~\rm{GeV} \,.
\end{equation}

\begin{figure}
  \centering

\begin{subfigure}
\centering
\includegraphics[width=0.65\textwidth]{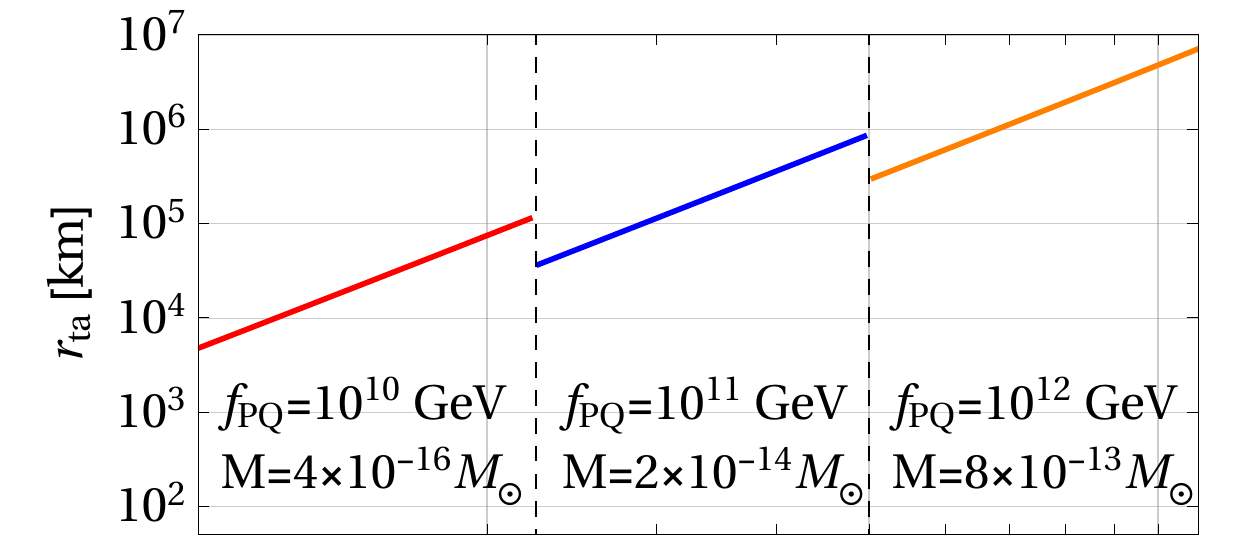}
\end{subfigure}
\begin{subfigure}
\centering
\includegraphics[width=0.65\textwidth]{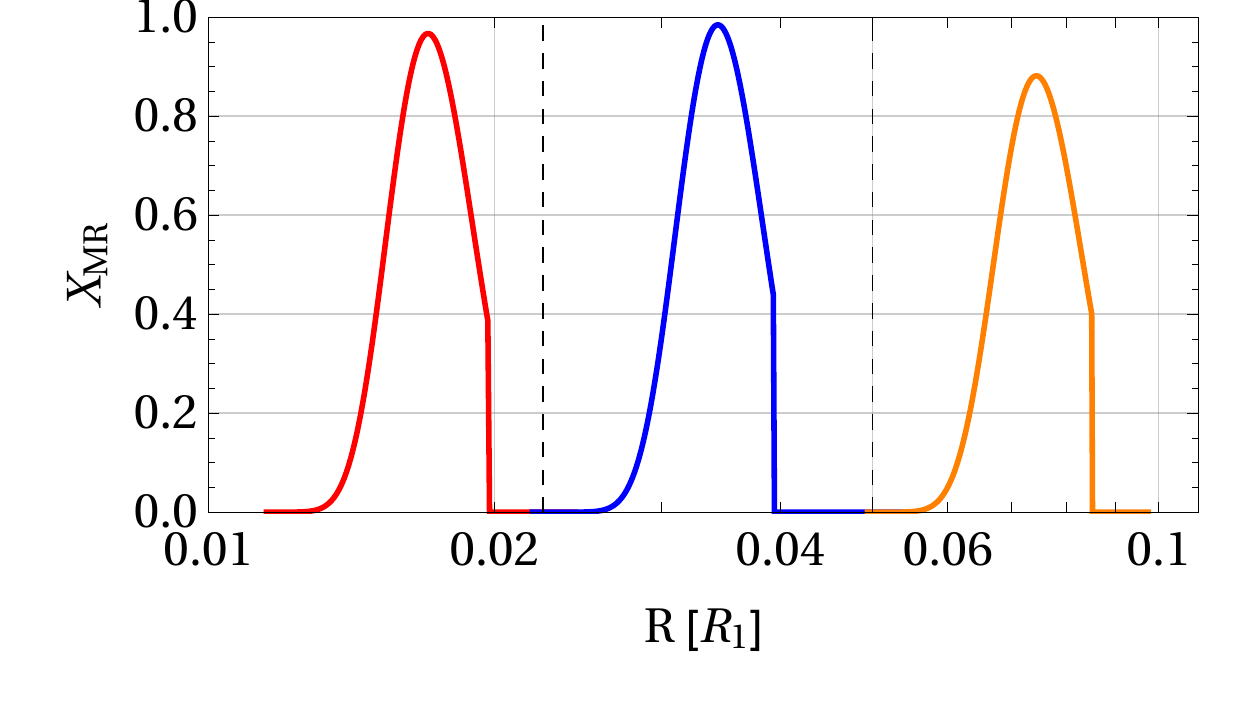}
\end{subfigure} 
 
\caption{{\it Upper panel:} The physical radius of an over-density with mass $M$ at the time when it decouples from the Hubble flow, as a function of its initial comoving size $R$, cf. \cref{eq:rta}. The masses are chosen such that they correspond to the peak values of the minicluster mass function, cf. \cref{tab:peak}. {\it Lower panel:} Slices through the dimensionless distribution function $X_{MR}$, defined in \cref{eq:XMR}, for constant $M$. Again the masses are fixed to the peak values of $X_{MR}$, as in the upper panel. For both panels the reference scale is the co-moving horizon size at 1~GeV, $R_1$. 
\label{fig:radius}}
\end{figure}

Those estimates are in good agreement with the numerical results for
$r_{\rm ta}$, which are shown in the upper panel of \cref{fig:radius}. We show the relationship between $R$ and $r_{\rm ta}$ for different
characteristic masses, corresponding to our three choices of $f_\PQ$.
The abundance of objects of the corresponding masses can be seen in
the lower panel, where we give slices through the 2-dimensional
distributions at the corresponding values of $M$.  At the peaks of the
distributions shown in the lower panel we find physical sizes at turn
around of $3.9\times 10^4$, $1.9\times 10^5$ and $1.5\times
10^6$~km. In \cref{tab:peak} we give also the interval in $r_{\rm ta}$
corresponding to the masses for which the mass function $X_M$ is
larger than 1\% of its maximum. For given $\fPQ$ the sizes of the
miniclusters at turn-around span approximately one order of
magnitude.

\section{Discussion and Summary}
\label{sec:summary}

In this paper we have developed a semi-analytic method to estimate the
distribution of axion miniclusters from the re-alignment
mechanism. Starting from an assumption on the statistical properties
of the axion field shortly before the QCD phase transition, we
calculate the resulting energy density power spectrum at the point
when all axions have become non-relativistic and the zero-temperature
axion mass is recovered, which happens around a temperature of
100~MeV. Departing from the power spectrum we use a spherical collapse
model and a modified Press \& Schechter approach to obtain the
distribution of gravitationally bound miniclusters in mass and size at
the point when they decouple from the Hubble flow and start to
collapse gravitationally. We find that for a given PQ breaking scale,
the masses of miniclusters contributing relevantly to the energy density
span more than two orders of magnitude, with peak values which are
about two orders of magnitude smaller than obtained from naive
estimates.

\subsection{Adopted assumptions and limitations}

Let us re-state here our most important assumptions and point
out the limitations of our approach. 
\begin{enumerate}
\item
  {\it Initial power spectrum of the axion field.} As our initial
  condition we assume statistical properties of the axion field before
  the QCD phase transition, motivated by the physics that field
  values in causally disconnected regions at that time should be
  uncorrelated. This is implemented in terms of the initial field
  power spectrum $P_\theta(q)$, and our default assumption is a
  Gaussian correlator with a characteristic scale set by the horizon
  at temperatures of a few GeV, see \cref{eq:auto-k,eq:K,eq:P-G}. In a
  more realistic approach, this power spectrum should be determined
  from the non-linear field evolution from the PQ scale down to the
  QCD scale including cosmic strings.

\item
  {\it Harmonic approximation of the potential.} In our work we assume
  a quadratic axion potential without axion self-interactions in order
  to obtain linear evolution equations and decoupling of Fourier
  modes. In this way we cannot capture non-linear effects in locations
  with field values close to $\pm\pi$, which indeed are to be expected
  in this scenario. Numerical simulations performed in
  \rcite{Kolb:1993zz, Kolb:1993hw} obtain very dense objects at those
  locations, which are not captured in our calculations. The open
  question remains of how likely those objects are and what their
  contribution to the total energy density is.
  
\item 
  {\it Neglecting axions from cosmic strings and domain walls.} This
  is related to the previous item, since the harmonic approximation
  largely neglects the periodic nature of the axion field, and
  therefore the contribution to the axion energy density due to the
  decay of cosmic strings and the domain wall network cannot be
  described. Numerical simulations \cite{Hiramatsu:2012gg,
    Kawasaki:2014sqa, Stadler} show that this contribution can be
  substantial. Furthermore one expects this distribution not to be
  homogeneous and it may contribute also significantly to the power
  spectrum. Our approach neglects those effects and includes only the
  part from the re-alignment mechanism. The separation into
  re-alignment and string decay parts is largely artificial since both
  should follow from the same physics described by the non-linear
  field evolution. In the interpretation of our results one has to
  keep in mind that it includes only part of the power and additional
  contributions are expected.

\item
  {\it Gaussianity of density fluctuations.} In our method to
  calculate the distribution of collapsed objects we assume that the
  density fluctuation distribution is Gaussian, with the variance
  determined by the power spectrum. The validity of this assumption is
  not obvious, due to order-one size of the fluctuations as well as
  non-linear effects mentioned in the items above. Some non-linearity
  can be implemented in principle in our formalism by considering
  higher-order correlation functions. But non-linear effects can lead
  to long non-Gaussian tails of the distribution, which can modify
  the mass function, particularly at large masses.  
\end{enumerate}

In view of those points, our results should be considered as a step
towards the goal of obtaining a complete understanding of the
minicluster distribution. It allows simple estimates and parameter
dependence studies under the stated limitations. Future work will be
dedicated to relaxing those assumptions.

\subsection{Outlook and comments on observational consequences}

An important open question is the subsequent evolution of the
minicluster after turn-around. The two extreme possibilities are that
either axions within the minicluster decohere and form a virialized
system of dust-like particles \cite{Tinyakov:2015cgg}, or the coherent
field configuration collapses and admits a stable solution of the
field equation under self-gravity, forming a so-called bose or axion
star \cite{Ruffini:1969qy, Liebling:2012fv}, whose ultimate fate is
currently under discussion, see e.g.,
\rcite{Levkov:2016rkk,Helfer:2016ljl,Eby:2016cnq,Eby:2017xrr}.  While
the investigation of the minicluster evolution after decoupling from
the Hubble flow is beyond the scope of this work, our results on the
relevant distribution of masses and sizes at turn-around provide
useful input for such considerations.

Clearly the further evolution and the fate of miniclusters during the
hierarchical formation of dark matter halos and the large scale
structures has important consequences for axion dark matter searches.
If we assume that a fraction $f_{\rm MC}$ of the total dark matter is
in form of clumps with mass $\sim 10^{-13} M_\odot$ their number
density in our galaxy would be $f_{\rm MC} \times 10^{-44}$~cm$^{-3}
\sim f_{\rm MC} \times 10^{-5}$/(1~AU)$^3 \sim f_{\rm MC}$/solar
system.  The flux on Earth would be $f_{\rm MC} \times 10^{-37}\,\rm
cm^{-2} \, s^{-1}$, and the frequency with which such a clump passes through a detector at Earth would be 
\begin{equation}
  \frac{f_{\rm MC} }{t_{\rm Univ}} \left(\frac{\text{clump size}}{10^6\,\rm km}\right)^2 
\end{equation}
with $t_{\rm Univ}$ being the age of the Universe. Hence the dark matter component bound in
such objects is invisible to axion haloscopes such as the experiments described in 
\rcite{Stern:2016bbw, TheMADMAXWorkingGroup:2016hpc} and their
expected event rate would be suppressed by a factor $(1-f_{\rm
  MC})$. On the other hand, if the final state of axion miniclusters
is only loosely bound, they might be tidally disrupted in the galaxy,
leading to potential signals in axion haloscopes
\cite{Tinyakov:2015cgg}.  The clumpy structure of dark matter halos
due to the presence of miniclusters may lead to observable signals in
femto-lensing \cite{Kolb:1995bu} or micro-lensing
\cite{Fairbairn:2017dmf,Fairbairn:2017sil}. Again an important
question to be answered in this context is about the size and masses
of those objects today.

\subsection*{Acknowledgements}

We thank Arthur Hebecker, J\"org J\"ackel and Javier Redondo for
useful discussions.  This project has received funding from the
European Union’s Horizon 2020 research and innovation programme under
the Marie Sklodowska-Curie grant agreement No 674896
(Elusives). A.P.\ acknowledges the support by the DFG-funded Doctoral
School KSETA.

\appendix
\section{Solving the equation of motion}
\label{app:EOM}

In this appendix we discuss how we solve the equation of motion of the
axion field. Since the Fourier modes evolve independently in the
harmonic approximation, we can make the ansatz
$\theta_k(a)=\theta_kf_k(a)$, \cref{eq:initial_cond}, with $a$ being
the scale factor. Then \cref{eq:eom-k} becomes an evolution equation
for $f_k$:
\begin{equation}\label{eq:eom-fk}
  \ddot f_k + 3H(T)\dot f_k + \omega_k^2(T) f_k = 0 \,,\qquad
\omega_k^2(T) = \frac{k^2}{a^2} + m(T)^2 \,.
\end{equation}
The temperature dependence of the Hubble rate $H=\dot{a}/a$ is determined by the Friedmann equation
\begin{equation}\label{eq:fried}
H^2=\frac{8\pi}{3M^2_{\rm{Pl}}}\rho \, ,
\end{equation}
where $M_{\rm{Pl}}$ is the Planck mass and $\rho$ is the energy density of the Universe. Since at the times we are interested in the Universe is radiation dominated, $\rho$ can be expressed as
\begin{equation}\label{eq:en-dens}
\rho=\frac{\pi^2}{30}g_{\rho}T^4 \, .
\end{equation}
The relativistic degrees of freedom $g_{\rho}$ depend on the temperature $T$ of the
Universe. In \rcite{Borsanyi:2016ksw} $g_{\rho}$ is determined by
lattice calculations in the relevant range of temperatures. We use the
tabulated values given therein and cubic spline interpolation to find $g_{\rho}(T)$. In
the evolution equation for $f_k$ not only the Hubble rate but also the
axion mass $m(T)^2=\chi(T)/\fPQ^2$ depends on the temperature. To find
$m(T)$ we use the tabulated results for the topological susceptibility
$\chi(T)$ of the lattice calculations in \rcite{Borsanyi:2016ksw} and
employ cubic spline interpolation. With $H(T)$ and $m(T)$ at hand, it
is convenient to rewrite the evolution equation in terms of
temperature being the independent variable instead of time. With the
initial condition $f_k(T_i)=1$ for all $k$, we can solve the resulting
equation numerically to find $f_k(T)$.

Soon after $T_{\rm{osc}}$, defined by
$3H(T_{\rm{osc}})=m(T_{\rm{osc}})$, rapid oscillations in $f_k$
commence since the mass term dominates the evolution equation and we
are dealing effectively with the equation of motion of an under-damped
oscillator. This motivates a WKB ansatz for $f_k$ of the form
\begin{equation}\label{eq:WKB-ansatz}
f_k=2A_k\cos\Phi_k \qquad (T<T_{\rm{osc}}) 
\end{equation}
with slowly varying amplitude $A_k$: $\dot{A}_k/A_k\ll \omega_k$, $\ddot{A}_k/A_k\ll \omega_k^2$.
In this approximation the evolution of $A_k$ respectively $\Phi_k$ is determined by
\begin{align}
\label{eq:a-evo}\dot{A}_k+\frac{1}{2}A_k\left(3H+\frac{\dot{\omega}_k}{\omega_k}\right)&=0 \, , \\
\label{eq:phi-evo}\dot{\Phi}^2-\omega^2_k&=0 \ .
\end{align}
The initial conditions for $A_k$ and $\Phi_k$ are found by matching the ansatz in \cref{eq:WKB-ansatz} and its derivative at a temperature $T_{\rm{WKB}}<T_{\rm{osc}}$ to the result of the full numerical solution of \eqref{eq:eom-fk}, i.e.,
\begin{align}
2A_k(T_{\rm{WKB}})\cos\Phi_k(T_{\rm{WKB}})=f_k(T_{\rm{WKB}})\, , \\
2A_k(T_{\rm{WKB}})\sin\Phi_k(T_{\rm{WKB}})=\frac{\dot{f}_k(T_{\rm{WKB}})}{\omega_k(T_{\rm{WKB}})} \ .
\end{align}
Supplemented with these initial conditions we can numerically solve
\cref{eq:a-evo} and \cref{eq:phi-evo} to find $f_k(T)$ at any
temperature $T$. For the matching temperature $T_{\rm{WKB}}$ between
the WKB ansatz and the full numerical result we chose
$T_{\rm{WKB}}=0.5 T_{\rm{osc}}$. We have checked that with this choice
the WKB approximation provides an excellent fit to a full numerical
solution. Note that for constant $g_\rho$ and constant axion mass,
\cref{eq:a-evo,eq:phi-evo} have a simple analytic solution. We decide
to solve them numerically to implement the full $T$ dependence of the
relativistic degrees of freedom and the axion mass down to $T_\star =
100$~MeV, where the power spectrum becomes constant.


\bibliographystyle{JHEP_improved}
\bibliography{bibliography}{}

\end{document}